
\input harvmac
                  \def\d{\delta}
    \def\e{\eta}        \def\f{\phi}       
             
\def\l{\lambda}   \def\L{\Lambda}             
\def\r{\rho}          \def\o{\omega}     
\def\p{\psi}              \def\s{\sigma}     
\def\th{\theta}     \def\t{\tau}       
              
\def\ep{\epsilon} \def\vep{\varepsilon}
%

   \def\CL{{\cal L}}

%
\def\rd{\partial}

\def\darr#1{\raise1.5ex\hbox{$\leftrightarrow$}\mkern-16.5mu #1}
\def\Ha{{1\over2}}

\def\fr#1#2{{\textstyle{#1\over#2}}}
\def\Fr#1#2{{#1 \over #2}}

\def\rf#1{\fr{\rd}{\rd #1}}

\def\Rf#1#2{\fr{\rd #1}{\rd #2}}


\def\slag{\Fr{i}{4}\l ^{ab}\,Tr\,\s_{ab}\,\L^{-1}\dot{\L}}

\def\spin{\Fr{1}{2}\l^{ab}\L\,\s_{ab}\,\L^{-1}}
\def\Df#1{\Fr{d}{d #1}}
%

\def\np#1#2#3{, Nucl.\ Phys.\ {{\bf #1}} {(#2)} {#3}.}
\def\pr#1#2#3{, Phys.\ Rev.\ {{\bf #1}} {(#2)} {#3}.}

\def\ap#1#2#3{, Ann.\ Phys.\ {{\bf #1}} {(#2)} {#3}.}
\def\jmp#1#2#3{, J.\ Math.\ Phys.\ {{\bf #1}} {(#2)} {#3}.}
\def\nc#1#2#3{, Il\ Nuovo\ Cimento\ {{\bf #1}} {(#2)} {#3}.}
\def\prs#1#2#3{, Proc.\ Roy.\ Soc.(London)\ {{\bf #1}} {(#2)} {#3}.}

\Title{\vbox{\baselineskip12pt\hbox{KAIST-CHEP-93/03}\hbox{YUMS-93-8}}}
{\vbox{\centerline{Relation between Classical and}
       \vskip2pt\centerline{Pseudo-classical Spinning Particle}}}

\centerline{Jin-Ho Cho\footnote{$^\dagger$}{(jycho@rcunix.kotel.co.kr)} $^1$,
Seungjoon Hyun\footnote{$^\ddagger$}{(sjh92@krysucc1.BITNET)} $^2$ and
Jae-Kwan Kim $^1$}
\bigskip\centerline{$^1$ Department of Physics}
\centerline{Korea Advanced Institute of Science and Technology}
\centerline{373-1 Yusung-ku, Taejon, 305-338, Korea}
\bigskip\centerline{$^2$ Institute for Mathematical Sciences}
\centerline{Yonsei University, Seoul, 120-749, Korea}

\vskip .3in
The spin degrees of freedom for the relativistic particle are described
 by either Lorentz group variables (classically) or Grassmann variables
(pseudo-classically). The relationship between those two descriptions are
given. In doing that, appropriate constraints are constructed to put
into the lagrangian. Especially a natural relation of Lorentz group
variables and
Grassmann variables is obtained. Hopf fibration relating the spin
components to the group is just the right transformation of the spin
components under Lorentz group. And with the relation just mentioned,
pseudo-classical lagrangian is derived naturally from the classical one.

\Date{} 

\baselineskip=20pt plus 2pt minus 2pt

\eject

Since the introduction of Clifford algebra by
Dirac\ref\rdi{P. A. M. Dirac\prs{A117}{1928}{610}},
several efforts have been made to understand the spin in the classical
sense and, at least, two methods have been developed. In one approach,
which we shall call `classical' approach, the
spin degrees of freedom are described by Lorentz group variables which
can be introduced in the lagrangian by either the requirement of
Poincar\'{e} invariance\ref\rha{A. Hanson, T. Regge and C. Teitelboim,
Accademia
Nazionale dei Lincei \hfil\break(Roma, 1976).} or Hopf
fibration\ref\rba{A. P. Balachandran, G. Marmo, B.-S. Skagerstam and
A. Stern,\ {\it Gauge Symmetries and Fiber
Bundles}, (Springer-Verlag, Berlin, 1983).} and the spin value is fixed
arbitrarily. (But only integer or half integer values are possible at the
quantum level\rba\ref\rgu{K. Gupta and A. Stern
\pr{D44}{1991}{2432} }.)

In the other, the spin degrees of freedom are described by Grassmann variables
which also may be put into the lagrangian through either the classical
analogue of Clifford algebra\ref\rbe{F. A. Berezin and
M. S. Marinov\ap{104}{1977}{336}}\ref\rbr{L. Brink, P. Di Vecchia
and P. Howe\np{B118}{1977}{76}} or
supersymmetry\ref\rca{A. Barducci, R. Casalbuoni and
L. Lusanna\nc{35A}{1976}{377}}
\nref\rhe{M. Henneaux and C. Teitelboim\ap{143}{1982}{127}}--
\ref\rpo{A. M. Polyakov,\ {\it Gauge Fields and Strings}, (Harwood academic
publishers, Chur, 1987).}. But in this case the spin is of zero
magnitude due to the Grassmann property, whereas it obtains non zero value
at the quantum level. So this latter case is especially called
`pseudo-classical'.

In these developments, we have found that, in principle, the
spin can be formulated classically and these give another good example for
the constrained dynamical system. Now, we have a natural question about the
relationship between those two formulations, as the answer to this
question is expected to give us a large amount of insight about spin.

This paper is an attempt to answer the question. The main idea is that
the antisymmetric property of the spin components can be realized by
Grassmann variables. The cost for this realization is that we lose direct
physical significance about the spin because its magnitude
comes to zero. But we do not have to worry about that. We can always restore
the nonzero magnitude through the quantization of the system. Therefore the
Grassmann variable formulation may find its significance in the
pseudo-classical sense.

In advance of the main story, for our convenience and notation fixing, it would
be
better to remind us those two `classical' and `pseudo-classical' formalisms
for the spinning particle.
Nonrelativistic spin comes in physics from the universal covering group
$SU(2)$ of $SO(3)$. This can be realized through quantum mechanics just
to understand Stern-Gerlach experiment. When it comes to quantum field
theory or any relativistic theory, it is natural to
extend $SO(3)$ to $SO(3,1)$. Considering the particle mass, we come to
Poincar\'{e} group $ISO(3,1)$, which has two Casimir invariants
(mass and spin).
But we have only a few choices for the Poincar\'{e} invariant
lagrangian\rha.
One simple and natural choice is as follows\rba\rgu.
\eqn\ela{\CL_{cl}\,=p^a \dot{x}_a+\Fr{i}{4}\l ^{ab}\,Tr\,\s_{ab}\
\L^{-1}\dot{\L},}
where $\l^{ab}$ is the component of a fixed element in $so(3,1)$
along the generator $\s_{ab}$ and $\L\,\in\,SO(3,1)$.

For the representation of
$(\s_{ab})^{cd}=\,-i\,(\d^c_a\,\d^d_b\,-\d^d_a\,\d^c_b)$, which satisfies
$[\,\s_{ab}\,,\s_{cd}\,]=-i\,(\s_{bc}\,\e_{ad}\,+\,\s_{ad}\,
\e_{bc}\,-\,\s_{bd}\,\e_{ac}\,-\,\s_{ac}\,\e_{bd})$, the
second term can be rewritten as
\eqn\els{\CL_{spin}\,=\,\slag\,=\,\Ha\l^{ab}\L_{cb}\dot{\L}^c{}_a.}

The spin components $S^{ab}$ are related to $\L$ as
\eqn\eho{\Ha\,S^{ab}\,\s_{ab}\,=\,\spin,}
which is a kind of Hopf projection reducing one degree of freedom of
$\L$ to get a constraint surface ( $S^2\,=\,\l^2$ ) for the spin components.

In fact, making use of the property of $\L\, ,\CL_{spin}$ can be redescribed as
\eqn\elsp{\CL_{spin}\,=\,\Ha\,\l^{cd}\,\L^a{}_c\L^b{}_d\L_a{}^e\dot{\L}_{be}
\,=\,\Ha\,S^{ab}(\L\,\dot{\L}^{-1})_{ab}}
and $\L\,\dot{\L}^{-1}$ is the relativistic analogue of angular
velocity\rha\ . Hence the spin components defined by Hopf projection
represent the right form of spin components transformed from some fixed
one $\l$ under $ISO(3,1)$.

The system has two implicit constraints:
\eqn\econ{p^2\,+\,m^2\,=0\,,\,\,\,S^2\,=\,\l^2,}
where the first equation comes from the transformation property of
momentum under $ISO(3,1)$
\eqn\emo{p^a\,=\,\L^a{}_b\,\r^b\,,\,\,\,\r^2\,=\,-m^2,}
and the second comes from, as mentioned, that of spin components under
$ISO(3,1)$, i.e.,Hopf projection.
\eqn\esmo{S^{ab}\,=\,\L^a{}_c\,\L^b{}_d\,\l^{cd}.}

In \rba, they gave another condition $S^{ab}p_b\,=\,0$, to pick specific
values of $\r$ and $\l$. But in our case, we do not interpret $S^{ab}$ as the
one which corresponds to nonrelativistic spin momentum. Such a physical
spin momentum can be defined by the projection of the conserved angular
momentum
along the direction normal to $p^a$\ref\rga{C. A. P. Galvao and C. Teitelboim
\jmp{21}{1980}{1863}}. But that specific procedure is not
necessary here.

The equations of motion can be derived in the standard way by the
variation of $x_a$ and $\L$:
\eqn\evx{\d\,x^a\,\,\,\Longrightarrow\,\,\,\dot{p}^a\,=\,0,}
\eqn\evl{\d\,\L\,=\,i\,\ep^{ab}\s_{ab}\,\L\,\,\,\Longrightarrow\,\,\,
\dot{J}^{ab}\,=\,\Df{\t}(\,x^ap^b-x^bp^a+S^{ab}\,)\,=\,0.}
Therefore the equations of motion describe the conservation of momentum and
angular momentum.

But there is another way for the description of the spinning particle;
classical mechanics can be generalized  to include new Grassmann dynamical
variables $\p^a$ which account for spin degrees of freedom\rbe.
Due to the anticommuting property of
Grassmann variables, the plausible form of the lagrangian for the massless
fermionic particle will be\rga
\eqn\sulag{\CL_{pcl}\,=\,p^a\dot{x}_a\,-\Ha\,\p^a\dot{\p}_a.}
But for the massive case, we get some hints from Dirac equation, which
will reflect correct constraints, as Klein-Gordon equation does. That is,
we get two constraints; $p^2\,+\,m^2\,=\,0$ from Klein-Gordon equation
and $p^a\p_a\,+\,m\,\p^*\,=\,0$ from Dirac equation\rga\ ($\p^*$ adjusts
the oddness of the second constraint). We take these as the first class
in passing to the classical limit. So the full lagrangian will be
\eqn\susy{\CL\,=\,p^a\dot{x}_a\,-\Ha\,\p^a\dot{\p}_a\,-\,\Ha\,\p^*\dot{\p}^*
\,-\,N\,(\,p^2\,+\,m^2\,)\,-\,M\,(\,p^a\p_a\,+\,m\,\p^*\,),}
where $N$ and $M$ are bosonic and fermionic lagrange multiplier,
respectively and the kinetic term for $\p^*$ is introduced to deal with
$\p^*$ dynamically. There is another route to get the equation \susy. At
first, the requirement of global supersymmetry gives us the lagrangian
\sulag. And the world line local supersymmetry leads us the very
lagrangian \susy\ ($\p^*$ shows up to localize the cosmological
term)\rpo. A third path will be shown in the next section.

The equations of motion can be obtained by varying
$p^a,\,x^a,\,\p^a,\,\p^*,\,N$ and $M$.
\eqn\evap{\d\,p^a\,\,\,\Longrightarrow\,\,\,\dot{x}^a\,-\,2Np^a\,-\,M\p^a\,=\,0,???}
\eqn\evax{\d\,x^a\,\,\,\Longrightarrow\,\,\,\dot{p}^a\,=\,0,}
\eqn\evp{\d\,\p^a\,\,\,\Longrightarrow\,\,\,\dot{\p}^a\,-\,Mp^a\,=\,0,}
\eqn\evpf{\d\,\p^*\,\,\,\Longrightarrow\,\,\,\dot{\p}^*\,-\,Mm\,=\,0,}
\eqn\evM{\d\,N\,\,\,\Longrightarrow\,\,\,p^2\,+\,m^2\,=\,0,}
\eqn\evN{\d\,M\,\,\,\Longrightarrow\,\,\,p^a\p_a\,+\,m\p^*\,=\,0.}

And Poincar\'{e} invariance generates conserved charges through Noether's
procedure. The Poincar\'{e} behavior of the dynamical variables are as
follows\rga.
\eqn\edx{\d\,x^a\,=\,\o^a{}_bx^b\,+\,\vep^a\,,}
\eqn\edp{\d\,p^a\,=\,\o^a{}_bp^b\,,}
\eqn\edps{\d\,\p^a\,=\,\o^a{}_b\p^b\,,}
\eqn\edpf{\d\,\p^*\,=\,0,}
with $\o_{ab}\,=\,-\,\o_{ba}$. The conserved charges are given also as
\eqn\ecp{\dot{p}^a\,=\,0\,\,\,,\,\,\,\dot{J}^{ab}\,=\,
\Df{\t}(\,x^ap^b-x^bp^a+\p^a\p^b\,)\,=\,0.}

Now we have two formalisms which describe the same
object, i.e.,the spinning particle. But they are different in their
formulations. For the description of spin degrees of freedom, one
considers Lorentz
group elements dynamically while the other does Grassmann variables. And
one deals with the constraints implicitly through Hopf projection
whereas the other does those things explicitly in the lagrangian.

In spite of these differences, they also have features in common. They are
Poincar\'{e} invariant with the same conserved charges and there are
two constraints concerned with the transformation properties of
momentum and spin components related with those conserved charges,
under Poincar\'{e} group.
And the two formalisms are of the first order for the spin variables. So
we come to think of the relation between them.

For that purpose we first put the implicit constraints into the
lagrangian in the first formalism. But we have a problem here.
The constraint
$p^2\,+\,m^2\,=\,0$ is of good form to put into lagrangian, whereas
$S^2\,-\,\l^2\,=\,0$ is not since $S^2$ vanishes at
the pseudo-classical level due to Grassmann property. And the
constraint $S^{ab}\,=\,\L^a{}_c\,\L^b{}_d\,\l^{cd}$ is not desirable  either
as it contains $\L$'s in the right hand side (we want such a form that
the variables are constrained only by some fixed values like $\l^{ab}$ or
$\r^a$).  The appropriate candidate may come from the composition of $p^a$
and $S^{ab}$ given by constraints,\emo\ and \esmo. Thus we use
$\L_a{}^k\,S^{ab}\,p_b\,-\,\l^{kb}\r_b\,=\,0$ to get the following term,
\eqn\enco{M_k\,(\L_a{}^k\,S^{ab}\,p_b\,-\,\l^{kb}\r_b\,)\,=\,0,}
where $M_k$'s are lagrange multipliers.
Now lagrangian is written as
\eqn\eccc{\CL\,=\,p^a\dot{x}_a\,-\,N\,(\,p^2\,+\,m^2\,)\,+
\,\Ha\,S^{ab}\L_{ac}\dot{\L}_b{}^c\,
-\,M_k\,(\L_a{}^kS^{ab}p_b\,-\,\l^{ka}\r_a).}

To get pseudo-classical lagrangian from this one, we make use of the
anticommuting property of spin components. We represent $\l^{ab}$ as
$\th^a\,\th^b$, where $\th$'s satisfy $\{\,\th^a\,,\th^b\,\}\,=\,0$. But
for this representation, we cost the value of $\l^2$. So this may be
called `pseudo-classical representation'.

We can also obtain dynamical variable $\f$ corresponding to $\L$ through
$\f^a\,=\,\L^a{}_b\,\th^b$. Then $\f$'s also satisfy
$\{\,\f^a\,,\f^b\,\}\,=\,0$, which can be checked easily. And we may
restore Lorentz element from $\f$ by differentiation, as the variable
$\f$ is linear in $\th$ and the coefficient is the element of Lorentz
group $SO(3,1)$ (see the appendix).

Now we rephrase the lagrangian in terms of $\f$. The spin components become
\eqn\echa{S^{ab}\,=\,\th^c\th^d\,\L^a{}_c\,\L^b{}_d\,=\,\f^a\,\f^b}
and the kinetic term for spin variable changes also as
\eqn\eka{\Ha\l^{ab}\L_{cb}\dot{\L}^c{}_a\,
=\,\Ha\,\th^a\th^b\L_{cb}\dot{\L}^c{}_a\,=\,-\Ha\,\f_c\dot{\f}^c.}
And the second constraint term becomes
\eqn\echc{M_k\,(\L_a{}^kS^{ab}p_b\,-\,\l^{ka}\r_a)\,
=\,M_k\,(\th^k\f^bp_b\,-\,\th^k\th^b\r_b)\,
=\,M\,(\f^ap_a\,-\,\th^a\r_a),}
where we set $M_k\,\th^k\,=\,M$. $\r$ can be written in terms of $m$
by transformation from the particle rest frame.
\eqn\etr{\r_b\,=\,-\tilde{\L}_b{}^{0}\,m.}
Then $-\th^a\r_a\,=\,\th^a\tilde{\L}_a{}^{0}m\,$  and we get another
variable $\f^*\,=\,\th^a\tilde{\L}_a{}^{0}$ anticommuting with the
other Grassmann variables, but it is not independent of the others (in
fact, $\f^0\f^1\f^2\f^3\f^*\,=\,0$). Therefore the classical lagrangian
becomes
\eqn\efff{\CL\,=\,p^a\dot{x}_a\,-\Ha\,\f^a\dot{\f}_a\,-\,\Ha\,\f^*\dot{\f}^*
\,-\,N\,(\,p^2\,+\,m^2\,)\,-\,M\,(\,p^a\f_a\,+\,m\,\f^*\,)\,
+\,v\f^0\f^1\f^2\f^3\f^*,}
where the new dynamical variable $\f^*$ is treated dynamically so that
the constraint, $p^a\f_a\,+\,m\,\f^*=0$, becomes the first class.
And the last term, accounting for its dependency on others, is included due
to the Grassmann nature.

The transformation properties of those Grassmann dynamical variables are
straightforwardly obtained as
\eqn\etran{\bar{\f}^a\,=\,\bar{\L}^a{}_b\th^b\,
=\,\L^{'a}{}_c\L^c{}_b\th^b\,=\,\L^{'a}{}_c\f^c,}
\eqn\etraf{\bar{\f}^*\,=\,\th^a\tilde{\L}_a{}^{0}\,=\,\f^*,}
where $\tilde{\L}$ denotes the transformation from the particle rest
frame to the initial frame where the $\r^a$ and $\l^{ab}$ are taken.
Now we see the $\f$'s have the same property as $\p$'s and hence they can
be identified. So the lagrangian \efff\  is exactly the same as the one
\eccc\ except the last constraint term on $\f^*$. But this term may be absorbed
into $M$ by redefinition, thus gives no extra effect at the classical level.
In the quantum case, the story becomes a little different. In odd dimension,
the quantum version of the corresponding constraint plays an important
role of projection operator to decrease the degrees of freedom of the
irreducible representation for the quantized algebra of Grassmann
variables (Clifford algebra), thus gives the correct dimensionality of
Dirac spinor \ref\rpl{M. S. Plyushchy, ``Pseudoclassical
Description of the Massive Spinning Particle in $d$ Dimension,'' IHEP
Preprint}\foot{where the same lagrangian is derived in another way.}.
In even dimension the projection operator becomes trivial and the
constraint is just the irreducibility condition for Clifford algebra.

Let us conclude this letter with some remarks and summary.
In this paper, the relation between classical and pseudo-classical
lagrangian for the spinning particle has been established through `the
pseudo-classical representation' of the initial spin components $\l^{ab}$.
Throughout that procedure, we have learned that

\item{1.} Lorentz group elements are deeply related with Grassmann elements to
change the spin variable pseudo-classically and

\item{2.}` Hopf projection' specifies nothing but the transformation property
of
spin components.

\item{3.} The `implicit' constraints of classical lagrangian have been made to
 be
`explicit', and

\item{4.} through that procedure $\p^*$ is introduced naturally. And we
learned that

\item{5.} so introduced $\p^*$ carries the information of the transformation
from
the particle rest frame at every instant to the `initial' frame.

\item{6.} We gave it dynamics with the specification of another constraint
$\p^{0}\p^{1}\p^{2}\p^{3}\p^{*}\,=\,0$, which turns out to do an
important role in odd dimension,i.e.,to give the pseudo-classical
description of massive spinning particle in such dimensions.

\vskip .2in

All these arguments are irrespective of the space-time dimension. We
hope this work to give some hints for the study of Poincar\'{e} group
manifold. And the study of supersymmetry through this formalism will be
very interesting. Those works are in progress.

\bigbreak\bigskip\bigskip\noindent{\bf Acknowledgments}

This work was supported in part by the Korean Science and Engineering
Foundation.

\bigbreak\bigskip\bigskip\noindent{\bf Appendix}

\proclaim Theorem. Let $\p^a\,=\,K^a{}_b\,\th^b,\,\,\,\th\,\in$ Grassmann. Then
$K^a{}_b\,K_a{}^c\,=\,\d^c_b$, i.e., $K\,\in\,SO(3,1)$ .
\par
\vskip .1in

\proclaim proof.

Since $\p^a$ is linear in $\th$, we have
$$\Rf{\p^a}{\th^b}\,\th^b\,=\,\p^a.$$

And
$$\p^a\,\p^b\,=\,K^a{}_c\,K^b{}_d\,\th^c\,\th^d\,
=\,-K^a{}_c\,K^b{}_d\,\th^d\,\th^c\,=\,-\p^b\,\p^a.$$

Therefore $\p_a\,\Rf{\p^a}{\th^b}\,\th^b\,=\,\p_a\,\p^a\,=\,0$ implies
$$\p_a\,\Rf{\p^a}{\th^b}\,=\,k\,\th_b.$$

Thus,
$$\rf{\th_a}(\p_c\,\Rf{\p^c}{\th^b})\,
=\,\Rf{\p_c}{\th_a}\,\Rf{\p^c}{\th^b}\,=\,k\d^a_b,$$

where the first equality holds because $\p^c$ is linear in $\th^b.$

We may redefine $\p^{'a}\,=\,\fr{\p^a}{\sqrt{k}}$ to get
$$\Rf{\p{'c}}{\th_a}\,\Rf{\p^{'c}}{\th^b}\,=\,K_c^{'a}\,K^{'c}{}_b\,
=\,\d^a_b.$$

Hence $K'\,\in\,SO(3,1)$.

\listrefs
\bye